\documentclass[a4paper,12pt]{article}

\usepackage{amssymb,latexsym,amsmath}

\usepackage{fullpage}
\usepackage{nicefrac}
\usepackage{mathrsfs}%different \cal using \mathscr
\usepackage{slashed}% for dslash
\usepackage{bbold}
\usepackage{booktabs}
\usepackage{color,colortbl}

\usepackage[pdftex]{graphicx}

\makeatletter \@addtoreset{equation}{section}
\makeatother

\usepackage{xspace}
\usepackage{bbm}
\usepackage{tensind}
\tensordelimiter{`}

\usepackage[hidelinks]{hyperref}

\makeatletter
\g@addto@macro\bfseries{\boldmath}
\makeatother
\begin{document}

\begin{titlepage}
	\thispagestyle{empty}

	\vspace*{25pt}

	\begin{center}

	   	{ \LARGE{\bf Covariant formulation of BPS black holes and\\[4mm]
		the scalar weak gravity conjecture}}

		\vspace{40pt}

		{Gianguido Dall'Agata and Matteo Morittu}

		\vspace{25pt}

		{
		{\it  Dipartimento di Fisica ``Galileo Galilei''\\
		Universit\`a di Padova, Via Marzolo 8, 35131 Padova, Italy}

		\vspace{15pt}

	    {\it   INFN, Sezione di Padova \\
		Via Marzolo 8, 35131 Padova, Italy}
				}

		\vspace{40pt}

		{ABSTRACT}

		\vspace{40pt}

		% \textbf{Compiled on \today\ at \currenttime}
	\end{center}

	\vspace{10pt}

In this note we analyze the BPS black hole equations in extended supergravities and we find two interesting relations involving first and second derivatives of combinations of the central charges. One relation is a new identity that solely relies on the geometric properties of the scalar manifolds of extended supergravity theories. The other relation is a generalization of a scalar weak gravity conjecture relation conjectured by Palti and uses properties of the underlying black hole solution. We also provide for the first time an explicit covariant construction of the BPS squared action for such solutions.

\end{titlepage}

\baselineskip 6 mm

\renewcommand{\arraystretch}{1.2}

% \tableofcontents

%%%%%%%%%%%%%%%%%%%%%%%%%%%%%%%%%%%%%%%%%%%%%%%%%%%%%%%%%%%%%%
%%%%%%%%%%%%%%%%%%%%%%%%%%%%%%%%%%%%%%%%%%%%%%%%%%%%%%%%%%%%%%

\definecolor{shaded}{gray}{.85}

\allowdisplaybreaks

\section{Introduction} % (fold)
\label{sec:introduction}

The analysis of the necessary conditions for a generic effective theory to be compatible with the existence of an underlying  quantum theory of gravity has led in recent years to the formulation of a number of conjectures giving constraints that allow to distinguish good models from those that are in the so-called swampland \cite{Vafa:2005ui}.

One of the first such conjectures is the Weak Gravity Conjecture (WGC) \cite{ArkaniHamed:2006dz}, which for a U(1) boson coupled to gravity states that there must always exist a charged particle with mass $m$ and charge $q$ such that $m \leq g\,q\,M_p$.
There is by now strong evidence that such conjecture is correct (see \cite{Palti:2019pca} for a review and for an extensive list of references) and it has been generalized in various directions.
One of the general lessons one learns from these analyses is that if gravity is required  to always be the weakest force then one can constrain effective theories in various ways.

An interesting generalization of the WGC is its application to forces mediated by light scalar fields and one can find various proposals in the literature \cite{Palti:2017elp}--\cite{Freivogel:2019mtr}.
The first formulation of a version of the WGC to scalar fields is due to Palti \cite{Palti:2017elp}, who considered particles whose masses $m$ depend on some light scalar $\phi$ by means of trilinear couplings $\partial_\phi m$.
In this case the conjecture states that $\left(\partial_\phi m\right)^2 \geq m^2/M_P^2$, so that the force mediated by $\phi$ is stronger than the gravitational force.
While this applies only to the WGC scalars whose mass is a function of $\phi$, it still can give constraints on effective theories, which may even be too strong with respect to expectations \cite{Palti:2017elp}.
Still, this idea has been pushed even further by Gonzalo and Ib\'a\~nez in \cite{Gonzalo:2019gjp}, where a strong version of the scalar WGC has been proposed.
The idea is that scalar self-interactions should always be stronger than gravity, for any scalar in the theory.
This was summarized by the inequality
\begin{equation}\label{Gonzalo}
	2 (V''')^2 - V'' V'''' \geq \frac{(V'')^2}{M_p^2},
\end{equation}
where primes are derivatives of the scalar potential $V$ with respect to the scalar in exam.

This conjecture is much stronger, because it applies to any scalar, including massive mediators, and results in very strong constraints on effective theory models containing scalars.
While equation (\ref{Gonzalo}) has nice implications and seems compatible also with the swampland distance conjecture \cite{Ibaneztalk,Font:2019cxq}, it mixes ingredients that are clearly long-range with others that are related to short-range interactions (like the quartic couplings).
Its derivation from first principles, even in simple situations, is therefore challenging.

A different bound involving cubic and quartic interactions has been suggested in a footnote of \cite{Palti:2017elp}, where it was noted that, in the context of N=2 supergravity theories, the masses of supersymmetric black holes have to fulfill an interesting relation, which follows from special geometry, the geometry underlying the scalar $\sigma$-model.

In $N=2$ supergravity the central charge satisfies the algebraic identity \cite{Bellucci:2006zzb,Palti:2017elp}
\begin{equation}\label{N2rela}
	g^{i\bar\jmath} D_i \overline{D}_{\bar\jmath} |Z|^2 = n_V\, |Z|^2 + g^{i\bar\jmath} D_i Z \overline{D}_{\bar\jmath} \overline{Z},
\end{equation}
where $n_V$ is the number of vector multiplets.
This identity follows rather easily from the application of special geometry identities \cite{Ceresole:1995jg}
\begin{equation}
	\overline{D}_{\bar \jmath} Z = 0, \qquad D_i \overline{D}_{\bar \jmath} Z =  g_{i \bar\jmath}\, Z.
\end{equation}
Based on this relation, in a footnote of \cite{Palti:2017elp} there is a proposal for a scalar WGC constraint of the form
\begin{equation}\label{Palticonj}
	n\, m^2 +  g^{ij} \partial_i m \partial_j m \leq \frac12\,g^{ij} D_i \partial_j (m^2),
\end{equation}
where $n$ is the number of scalar fields coupling to the WGC state.
This is also a relation between mass, three-point and four-point couplings of the WGC states to scalar fields, but very different from (\ref{Gonzalo}).

In this note we want to give a stronger basis to a scalar WGC relation like (\ref{Palticonj}) by analyzing what happens for $N>2$ theories, where the central charge matrix has $N(N-1)/2$ entries and the supersymmetric black hole mass is equal to the largest of its eigenvalues 
\begin{equation}\label{ADM}
	M_{ADM} = |Z_1| > |Z_2| > \ldots> |Z_{N/2}|,
\end{equation}
where $Z_1$, $\ldots$, $Z_{N/2}$ are the eigenvalues of the central charge antisymmetric matrix $Z_{AB}$, written in normal form \cite{Ferrara:1980ra}.

Before generalizing (\ref{N2rela}), one should note that if we want to interpret it as a bound on the black hole mass we should rewrite it fully in terms of $M_{ADM} = |Z|$.
In this case the relation above can be expressed as
\begin{equation}\label{secondversion}
	D_i \overline{D}^i (|Z|^2) = 4 \, \partial_i |Z| \overline{\partial}^i |Z| + n_V |Z|^2.
\end{equation}
It is interesting to note the factor in front of the first derivative terms, which is going to be crucial in the correct identification of the generalization of such identity.

In this note we will prove two distinct relations.
The first is a purely algebraic identity, valid for any number of supersymmetries and reduces to (\ref{N2rela}) for $N=2$:
\begin{equation}\label{result1}
	D_a \overline{D}^a \left(Z_{AB}Z^{AB}\right) = D_a Z_{AB} \overline{D}^a Z^{AB} + n \, Z_{AB}Z^{AB},
\end{equation}
where
\begin{equation}
	n = n_V + \frac{(N-2)(N-3)}{2}
\end{equation}
and we used flat complex indices for the scalar derivatives.
This clearly reduces to (\ref{N2rela}) for $N=2$ and depends only on gravity multiplet scalars for $N>4$, as expected.
The interesting aspect is that the number $n$ corresponds precisely to half of the rank of the Hessian matrix of the black hole potential at ``fixed scalars'', therefore giving credit to the fact that in the relation between the mass and the three and four-point couplings only \emph{active scalars} should appear, where by active we mean scalars that support the black hole solutions and are not moduli.

As mentioned above, this relation is not suitable to be interpreted as a form of scalar WGC because the various derived quantities in (\ref{result1}) cannot be identified with the (square of) the ADM mass (\ref{ADM}).
We therefore analyzed more in detail the black hole solutions for $N>2$ and found that there is also a general differential relation on the ADM mass of such black holes, which uses some insights from the black hole solution.
This is going to be the generalization of (\ref{secondversion}) to an arbitrary number of supersymmetries and coincides with (\ref{result1}) for $N=2$.
This second relation is 
\begin{equation}\label{result2}
 	P^a{}_b D_a \overline{D}^b W^2 = 4 \, D_a W \overline{D}^a W + n \, W^2,
\end{equation}
where 
\begin{equation}
	W = \sqrt{\frac12\, Z^{AB} P_A{}^C P_B{}^D Z_{CD}}
\end{equation}
is the superpotential that can be identified with the ADM mass for BPS black holes in extended theories, $P^a{}_b$ is a projector on the space of active complex scalars and $P^A{}_B$ is a projector on the $R$-symmetry vector space to the bidimensional eigenspace related to the largest central charge value, according to (\ref{ADM}).

While deriving this last identity, we also work out a fully covariant formulation of the BPS equations and of the BPS squaring of the reduced action on the black hole solution. Since this has a general value for analyzing BPS black hole solutions in extended supergravities we provide explicitly this construction for $N=3$ and $N=4$ theories.

We then conclude with some comments on the physics of (\ref{result2}) and its compatibility and relation to the swampland distance conjecture.

% section introduction (end)

\section{Preliminaries} % (fold)
\label{sec:preliminaries}

When considering $N>2$ supergravity theories one should note and use the fact that the scalar $\sigma$-model is described by a homogeneous manifold $G/H$ of restricted type, because $H$ must contain the R-symmetry group U($N$) (SU(8) for $N=8$).
Also, duality invariance in 4 dimensions imposes that $G \subset Sp(2n_V,{\mathbb R})$, where $n_V$ is the total number of vector fields in the theory.
These facts allow us to perform a rather general analysis by considering the general structure of homogeneous manifolds and declining the various formulas to specific $N$ when necessary.
For the sake of self-consistency of this work, we recall here some preliminary relations already presented in \cite{Andrianopoli:1996ve,Andrianopoli:1996wf,Andrianopoli:1997pn}, whose conventions we mostly follow.

In order to parameterize the scalar manifold, we choose a coset representative $L$ in a basis that makes manifest duality relations.
We therefore take $L \in USp(n_V,n_V)$, i.e.~satisfying $L^\dagger \eta L = \eta = {\rm diag}\{{\mathbb 1}_{n_V}, -{\mathbb 1}_{n_V}\}$ and $L^T \Omega L = \Omega$, where $\Omega = \left(\begin{array}{cc}
0 & {\mathbb 1}_{n_V} \\ -{\mathbb 1}_{n_V} & 0
\end{array}\right)$.
A generic parameterization, useful in the following, is
\begin{equation}
	L = \frac{1}{\sqrt2}\left(\begin{array}{cc}
		f+ i\, h & f^* +i\, h^* \\	f-i\,h & f^*-i \, h^*
	\end{array}\right),
\end{equation}
where
\begin{eqnarray}
	&&f^T h = h^T f, \\[2mm]
	&&i (f^T h^* - h^T f^*) = - {\mathbb 1}.
\end{eqnarray}

Maurer--Cartan equations define the generic structure of the coset by producing its vielbeins and connection as
\begin{equation}
	{\cal W} = L^{-1} dL = \left(\begin{array}{cc}
		\omega & P^* \\ P & \omega^*
	\end{array}\right),
\end{equation}
which leads to the definitions
\begin{eqnarray}
	\omega &=& i( f^\dagger dh - h^\dagger df), \\[2mm]
	P &=& i (h^T df - f^T dh),
\end{eqnarray}
and to the relations 
\begin{eqnarray}
	d\omega + \omega \wedge \omega &=& P \wedge P^*, \\[2mm]
	DP &=& dP + \omega^* \wedge P + P \wedge \omega = 0.
\end{eqnarray}
We can make everything explicit by introducing flat indices on the coset manifold $G/H$.
Since $H =$ (S)U($N$) $\times H'$, we can write flat indices using a multi-index structure, combining U($N$) indices $A,B=1,\ldots,N$ and $H'$ indices $I,J =1,\ldots,n_h$, where $n_h$ is the dimension of the fundamental representation of $H'$.
More in detail, we split the real symplectic vector representation\footnote{The real embedding $G \subset Sp(2n_V,{\mathbb R})$ is appropriate for the explicit action of the duality group on the vector field strengths, while the complex embedding in USp($n_V,n_V$) is useful to write down the fermion transformation laws.} as $V^M = (V^\Lambda, V_\Lambda)$, $\Lambda = 1,\ldots,n_V$, and use the transformation properties of $L$ under the right action of $H$ to split the same vector in terms of a twofold complex tensor representation of (S)U($N$) and $H'$.
This means that the generic coset representative can be split accordingly, so that 
\begin{equation}
	\begin{split}
			f &= (f^\Lambda{}_{AB}, f^\Lambda{}_I), \\[2mm]
			h &= (h_{\Lambda AB}, h_{\Lambda I}),
	\end{split}
\end{equation}
and
\begin{equation}
	\begin{split}
			f^* &= (f^{\Lambda AB}, f^{\Lambda I}), \\[2mm]
			h^* &= (h_{\Lambda}{}^{AB}, h_{\Lambda}{}^I).
	\end{split}
\end{equation}
By using this decomposition we find
\begin{eqnarray}
	P_{AB I} &=& P_{I AB} = i(h_{\Lambda AB} d f^\Lambda{}_I - f^\Lambda{}_{AB}dh_{\Lambda I}), \\[2mm]
	P_{IJ} &=& i(h_{\Lambda I} d f^\Lambda{}_J - f^\Lambda{}_{I} dh_{\Lambda J}), \\[2mm]
	P_{ABCD} &=& i(h_{\Lambda AB} d f^\Lambda{}_{CD} - f^\Lambda{}_{AB}dh_{\Lambda CD})
\end{eqnarray}
and $P^{IAB} = (P_{IAB})^*$, $P^{IJ} = (P_{IJ})^*$ and $P^{ABCD} = (P_{ABCD})^*$.
Clearly such 1-forms correspond to vielbeins of $G/H$ in different ways according to the number of supersymmetries $N$.

{}For $N=3$, the scalar manifold is $G/H = {\rm SU}(3,n_V)/\left[{\rm SU}(3)\times {\rm SU}(n_V) \times U(1)\right]$, which has dimension $3\,n_V$.
This means that the flat vielbein indices lie in the $(3,n_V)$ representation of $H$ and hence $P_{ABCD} = P_{IJ} = 0$.

{}For $N=4$ the scalar manifold is $G/H =$ SU(1,1)/U(1) $\times$ SO(6,$n_V$)/[SO(6)$\times$ SO($n_V$)] and therefore the vielbein splits in two, $P_p$ being the complex vielbein of the first factor and $P_{IAB}$ in the $(6,n_V)$ representation of SU(4) $\times$ SO($n_V$) the complex vielbein of the second factor.
This implies $P_{ABCD} = \epsilon_{ABCD} P_p$, $P_{IJ} = \delta_{IJ} \overline{P}_{\bar{p}}$.
Moreover one should note that there is a complex self-duality condition on the vielbeins so that
\begin{equation}
	P_{IAB} = \frac12 \delta_{IJ} \, \epsilon_{ABCD} P^{JCD} = (P^{IAB})^*.
\end{equation}

{}For $N=5,6$ and 8 there are no vector multiplets and the scalar manifolds are SU(1,5)/U(5), SO$^*$(12)/U(6) and E$_{7(7)}$/SU(8), respectively of dimension 10, 30 and 70.
The vielbeins lie in the \textbf{5}, \textbf{15} and \textbf{35} representations of U(5), U(6) and SU(8) and are therefore always described by the complex $P_{ABCD}$.
However, the vector fields are in the \textbf{10}, \textbf{15}+\textbf{1} and \textbf{28} dimensional representations of their respective R-symmetry groups. 
This means that in the $N=6$ case there is a vector field that behaves as a matter vector field, being a singlet of the $R$-symmetry group.
We therefore have $P_{IJ} = 0$ and $P_{IAB} = 0$ for $N=5,8$, while for $N=6$ we also have $P_{\cdot AB} = \frac{1}{4!} \epsilon_{ABCDEF} P^{CDEF}$, where the $\cdot$ stands for the U(6) singlet.
Finally, in the $N=8$ case we also have a complex self-duality condition of the form
\begin{equation}
	P_{ABCD} = \frac{1}{4!} \epsilon_{ABCDEFGH} P^{EFGH}.
\end{equation}

{}From the relation $dL = L {\cal W}$ we can now obtain general relations for the covariant derivatives of the coset representatives:
\begin{equation}\label{Df}
	\begin{split}
	D f^\Lambda{}_{AB} &= f^{\Lambda I} P_{I AB} + \frac12\, f^{\Lambda CD} P_{CD AB}, \\[2mm]
	D f^\Lambda{}_{I} &= \frac12\,f^{\Lambda CD} P_{I CD} +  f^{\Lambda J} P_{J I}, 
	\end{split}
\end{equation}
where we also used that $f^* = (f^{\Lambda AB}, f^{\Lambda I})$.

In the following we are interested in relations that involve derivatives of the central charges of $N$-extended supergravities, for $N>2$.
Central charges are introduced as a symplectic product of a charge vector $Q = (p^\Lambda, q_{\Lambda})$ and the section vector ${\cal V} = (f^{\Lambda}, h_{\Lambda})$.
We therefore see that we have two types of charges
\begin{eqnarray}\label{ZAB}
	Z_{AB} &=& p^\Lambda h_{\Lambda AB} - q_{\Lambda} f^{\Lambda}{}_{AB},\\[2mm]
	\label{ZI}
	Z_I &=& p^\Lambda h_{\Lambda I} - q_{\Lambda} f^{\Lambda}{}_{I}.
\end{eqnarray}
The first set $Z_{AB}$ defines the actual central charges associated to the $N(N-1)/2$ graviphotons in the theory, while $Z_I$ are the matter charges, related to the possible additional vector multiplets (with the exception of the $N=6$ theory, as mentioned above).
It is then straightforward to obtain relations between these charges by taking their covariant derivatives, using (\ref{Df}), (\ref{ZAB}) and (\ref{ZI}):
\begin{eqnarray}\label{nablaZAB}
	D Z_{AB} &=& Z^I P_{I AB} + \frac12 \, Z^{CD} P_{CDAB}, \\[2mm]
	\label{nablaZI}
	D Z_I &=& Z^J P_{JI} + \frac12 \, Z^{CD} P_{I CD}.
\end{eqnarray}

In order to compute (second) derivatives of the central charges and of the ADM mass, we need the explicit expression of the derivatives we can obtain from (\ref{nablaZAB}) when projecting on the scalar $\sigma$-model vielbeins.
The exercise is straightforward and we report here the outcome for the different values of $N$:
\begin{eqnarray}\label{DN3}
	&&\begin{array}{ccll}
	N=3:& \phantom{space}& \displaystyle D = \frac12 P_{IAB} D^{IAB} + \frac12 P^{IAB} D_{IAB},  & \\[2mm]
	& & D^{I CD} Z_{AB} = 2\,\delta^{CD}_{AB} Z^{I}, & D_{I CD} Z_{AB} = 0,\\[2mm]
	& & D^{I CD} Z_J = \delta^I_J\, Z^{CD}, & D_{I CD} Z_J = 0.
	\end{array}\\[5mm]
\label{DN4}	&&\begin{array}{ccll}
	N=4: & \phantom{space}& \displaystyle D = \frac14 P_{IAB} D^{IAB} + \frac14 P^{IAB} D_{IAB}  &+ P_p D_p + 
	\overline{P}_{\bar p} D_{\bar p}, \\[2mm]
	&&D_{I CD} Z_{AB} = \epsilon_{ABCD}\, \delta_{IJ}\, Z^{J}, & D^{I CD} Z_{AB} = 2\,\delta^{CD}_{AB} Z^{I}, \\[2mm]
	&&D_{JAB} Z_I = \displaystyle\frac12\, \delta_{IJ}\, \epsilon_{ABCD}\, Z^{CD}, & D^{JAB} Z_{I} = \delta_I^J Z^{AB}, \\[2mm]
	&&D_{p} Z_{AB} = \displaystyle\frac12 \epsilon_{ABCD} Z^{CD}, & D_{\bar p} Z_{AB} = 0, \\[2mm]
	&&D_p Z_I = 0, & D_{\bar p} Z_{I} = \delta_{IJ} Z^J.
	\end{array}\\[5mm]
\label{DN5}	&&\begin{array}{ccl}
	N=5: & \phantom{space}& \displaystyle D = \frac{1}{4!} P_{ABCD} D^{ABCD} +\frac{1}{4!} P^{ABCD} D_{ABCD},  \\[5mm]
	&&D^{ABCD} Z_{EF} = 12 \,\delta_{EF}^{[AB}Z^{CD]}, \quad \qquad D_{ABCD} Z_{EF} = 0.
	\end{array}\\[5mm]
\label{DN6}	&&\begin{array}{ccl}
	N=6: & \phantom{space}& \displaystyle D = \frac{1}{4!} P_{ABCD} D^{ABCD} +\frac{1}{4!} P^{ABCD} D_{ABCD},    \\[5mm]
	&& D_{ABCD} Z_{EF} = \epsilon_{ABCDEF} \bar{Z},  \qquad D^{ABCD} \displaystyle Z_{EF} = \frac{4!}{2} \delta^{[AB}_{EF} Z^{CD]}, \\[2mm]
	&&\displaystyle D_{ABCD} Z = \frac12 \epsilon_{ABCDEF} Z^{EF},  \quad D^{ABCD} Z = 0.
	\end{array}\\[5mm]
\label{DN8}	&&\begin{array}{ccl}
	N=8: & \phantom{space}& \displaystyle D = \frac12\frac{1}{4!} P_{ABCD} D^{ABCD} +\frac12\frac{1}{4!} P^{ABCD} D_{ABCD} ,   \\[5mm]
	&&  \displaystyle D_{ABCD} Z_{EF} = \frac12\,\epsilon_{ABCDEFGH} Z^{GH}.  \quad D^{ABCD} Z_{EF} = 12 \,\delta^{[AB}_{EF} Z^{CD]}.
	\end{array}
\end{eqnarray}

% section preliminaries (end)

\section{The identity} % (fold)
\label{sec:the_identity}

In this section we provide the details of the derivation of the general algebraic identity (\ref{result1}).
The formula encompasses the specific forms we obtained for similar calculations done for different numbers of supersymmetries.
We therefore perform our calculations by using the derivative relations on the central charges obtained in the previous section, declined for specific $N$ in (\ref{DN3})--(\ref{DN8}), and applying them to the square of the central charges $Z_{AB} Z^{AB}$, which is an $H$-invariant tensor.

\paragraph{N=3 identity.}

The computation of the second derivative of the sum of the squares of the central charges can be easily obtained by applying the rules described in (\ref{DN3}) and leads directly to the desired result:
\begin{equation}\label{N3identity}
	\frac12 D^{I CD} D_{I CD}(Z_{AB} Z^{AB}) = \frac12 D^{I CD} Z_{AB} D_{I CD} Z^{AB}  + n_V\, Z_{AB} Z^{AB}.
\end{equation}

\paragraph{N=4 identity.} In the $N=4$ case, one has to be more careful because there are two factors in the $\sigma$-model and there is a duality constraint between $P_{IAB}$ and $P^{IAB}$.
This is also reflected in the numerical factors needed to obtain the correct result:
\begin{equation}
	\begin{split}
	&\frac14 D^{I CD} D_{I CD}\left(Z_{AB} Z^{AB}\right) + D_{p}D_{\bar p}\left(Z_{AB} Z^{AB}\right)= \\[2mm]
	=&\frac14 D^{I CD} Z_{AB} D_{I CD} Z^{AB} +\frac14 D_{I CD} Z_{AB} D^{I CD} Z^{AB} + D_{p} Z_{AB} D_{\bar p} Z^{AB}   \\[2mm]
	& +(1+n_V)  Z_{AB} Z^{AB},
	\end{split}
\end{equation}
where we identify $1=(N-2)(N-3)/2$.

\paragraph{$N=5$ identity.} In this case the identity follows again straightforwardly from the application of (\ref{DN5})
\begin{equation}
	 \frac{1}{4!} D_{CDEF} D^{CDEF} \left(Z_{AB} Z^{AB}\right) = \frac{1}{4!} D_{CDEF} Z^{AB} D^{CDEF} Z_{AB} + 3\,Z^{AB} {Z}_{AB},
\end{equation}
where we identify $3= (N-2)(N-3)/2$.

\paragraph{$N=6$ identity.} While the final relation in the $N=6$ case has the same structure as the previous ones, the derivation is a bit more delicate, because there is a vector in the gravity multiplet that is a singlet of the $R$-symmetry group and therefore its central charge behaves as a matter charge.
Anyway, by repeatedly using (\ref{DN6}) one obtains
\begin{equation}
	 \frac{1}{4!} D_{CDEF} D^{CDEF} \left(Z_{AB} Z^{AB}\right) = \frac{1}{4!} D_{CDEF} Z^{AB} D^{CDEF} Z_{AB} + 7\,Z^{AB} {Z}_{AB},
\end{equation}
where we identify $7 = 1+ (N-2)(N-3)/2$ and the extra unity corresponds to the vector that acts as a matter multiplet.

\paragraph{$N=8$ identity.} The only delicate point is once more the duality relation between the vielbeins.
This is the reason for the different coefficient in the formula with respect to the $N=6$ case.
Using (\ref{DN8}) we obtain
\begin{equation}
	\begin{array}{l}
	   	 \displaystyle \frac12 \,\frac{1}{4!} D_{CDEF} D^{CDEF}\left(Z_{AB} Z^{AB}\right) = \\[2mm]
		\displaystyle \frac12 \frac{1}{4!} D_{CDEF} Z^{AB} D^{CDEF} Z_{AB}+  \frac12 \frac{1}{4!} D_{CDEF} Z_{AB} D^{CDEF} Z^{AB} + 15\,Z^{AB} {Z}_{AB},
	\end{array}
\end{equation}
where we identify $15 = (N-2)(N-3)/2$.

\bigskip

\paragraph{General Form.} Altogether we can summarize all these identities in a single formula, where we use a single-index complex notation for the scalar fields:
\begin{equation}
	D_a \overline{D}^a \left(Z_{AB}Z^{AB}\right) = D_a Z_{AB} \overline{D}^a Z^{AB} + n \, Z_{AB}Z^{AB},
\end{equation}
where
\begin{equation}
	n = n_V + \frac{(N-2)(N-3)}{2}.
\end{equation}

As noted in the introduction, the number $n$ corresponds to half of the rank of the Hessian matrix of the black hole potentials at fixed scalars, but our derivation was fully general and did not make use of the black hole solution at any stage.
It is indeed an identity that follows by purely algebraic relations imposed by the geometry of the scalar $\sigma$-model.

% section the_identity (end)

\section{BPS black holes in $N=3$ supergravity} % (fold)
\label{sec:bps_black_holes_in_n_3_supergravity}

The identity derived in the previous section has general validity and reduces to the $N=2$ identity noted in \cite{Palti:2017elp} to argue that there may be a scalar WGC constraining cubic and quartic interactions.
However, for $N>2$ the combination $Z_{AB} Z^{AB}$ cannot be identified directly with the ADM mass and the first-derivative terms do not act on duality-invariant quantities, but directly on the central charges, hence giving expressions that depend on the basis.

For this reason we now analyze in detail the BPS rewriting of the reduced action of $N$-extended supergravity and propose a new relation that generalizes (\ref{secondversion}) for arbitrary $N$.

The general metric ansatz for an extremal, asymptotically flat black hole solution \cite{DallAgata:2011zkh} depends on a unique unknown function:
\begin{equation}
	ds^2 = -e^{2U(r)} dt^2 + e^{-2U(r)}dr^2 + r^2 d \Omega^2,
\end{equation}
where $d \Omega^2 = d \theta^2 + \sin^2 \theta\, d \phi^2$ is the line-element of a two-sphere and $U$ is the warp factor, which depends only on the radial variable to respect spherical symmetry.
The vector and scalar fields also satisfy the same spherical symmetry requirement, with electric and magnetic charges located at $r=0$.
We can therefore reduce the 4-dimensional supergravity action to a 1-dimensional action depending only on the $r$ variable, denoting derivatives with respect to $r$ by a prime.

In the case of $N=3$ supergravity \cite{Castellani:1985ka}, the reduced lagrangian is
\begin{equation}\label{LN3}
	{\cal L} = \frac12 \, P'_{IAB}P^{\prime IAB} + (U')^2 + e^{2U} V_{BH},
\end{equation}
where \cite{Andrianopoli:1996ve}
\begin{equation}
	V_{BH} = \frac12\, Z_{AB}Z^{AB} + Z_I Z^I.
\end{equation}
The BPS equations \cite{Andrianopoli:1996ve} follow from requiring the vanishing of the supersymmetry transformation of the fermions on this background:
\begin{eqnarray}
	&& \varepsilon_{A}' -\frac{i}{2} e^U \gamma^0 Z_{AB} \varepsilon^B = 0, \\[2mm]
	&& U' \varepsilon_{A} -i e^U \gamma^0 Z_{AB} \varepsilon^B = 0, \label{Uprime}\\[2mm]
	&& Z_{AB} \varepsilon_C \,\epsilon^{ABC} = 0, \label{Nred}\\[2mm]
	&& P^\prime_{IAB} \varepsilon_C \,\epsilon^{ABC} = 0, \\[2mm]
	&& P^\prime_{IAB} \varepsilon^B = i e^U  \, Z_I \gamma^0 \varepsilon_A.
\end{eqnarray}
The interpretation of these equations is that the first fixes the radial dependence of the supersymmetry spinor, the second gives the flow of the warp factor, the third projects away one component of the spinor, the fourth constrains the number of scalars flowing and finally the last one gives the flow equations of the scalar fields.
Essentially, we have first a reduction from $N=3$ to $N=2$ because of (\ref{Nred}) and then we recover the same type of equations as for the $N=2$ case, with the addition of a constraint on the active scalars.
To see this in detail, we define the normalized vector
\begin{equation}
	V_A \equiv \left(2 Z_{EF}Z^{EF}\right)^{-1/2}\, \epsilon_{ABC} Z^{BC},
\end{equation}
which is going to give the direction orthogonal to the preserved supersymmetry, according to (\ref{Nred}), and we use it to define the projector to its orthogonal subspace:
\begin{equation}
	P^A{}_B = \delta^A_B - V^A V_B.
\end{equation}
The correct set of BPS equations follows now as gradient flows activated by the superpotential
\begin{equation}\label{W}
	W = \sqrt{\frac12 \,Z_{CD} P^C{}_A P^D{}_B Z^{AB}},
\end{equation}
which coincides with the ADM mass of the solution.
We emphasize this definition of the superpotential, because it is going to be the expression that will be generalized to arbitrary $N$.

The first thing to note is that in this special instance ($N=3$) the superpotential reduces to
\begin{equation}\label{WN3}
	W = \sqrt{\frac12 Z_{AB} Z^{AB}},
\end{equation}
because the central charge is automatically orthogonal to the $V$ vector:
\begin{equation}
	Z^{AC}V_C \sim \epsilon_{CDE} Z^{AC} Z^{DE} = \epsilon_{CDE} Z^{A[C} Z^{DE]} = \epsilon_{CDE} Z^{[AC} Z^{DE]} = 0.
\end{equation}
In order to derive bosonic flow equations, we then have to impose two projectors on the Killing spinors to reduce supersymmetry to $N=1$ along the solution.
One projection follows straightforwardly from (\ref{Nred}), the other can be read from the (\ref{Uprime}) equation and is needed to relate the action of the $\gamma^0$ matrix on the spinor with the action of the central charge matrix:
\begin{eqnarray}
	&&i \, \gamma^0 \varepsilon^A = \frac{1}{W} \, Z^{AB} \varepsilon_B, \\[2mm]
	&&V^A \varepsilon_A = 0 \quad \Leftrightarrow \quad P_A^B \varepsilon_B = \varepsilon_A. \label{proj2N3}
\end{eqnarray}
Consistency of these projection operations is easy to check.
For instance
\begin{equation}
	(\gamma^0)^2 \varepsilon_A = i \,\frac{Z_{AB}}{W} \gamma^0 \varepsilon^B = \frac{1}{W^2} Z^{BC} Z_{AB} \varepsilon_C =\frac{1}{W^2}\left(W \epsilon^{BCD} V_D\, W\, \epsilon_{ABE} V^E\right) \varepsilon_C = -P^C{}_A \varepsilon_C,
\end{equation}
which correctly produces $(\gamma^0)^2 \varepsilon_A=-\varepsilon_A $ once (\ref{proj2N3}) is employed.
We see that, in addition to the equation fixing the Killing spinor, the 1/3 BPS black hole solution is determined by the following two BPS equations:
\begin{eqnarray}\label{flow31}
	U' &=& - \,e^U W, \\[2mm]
	P_{IAB}^\prime &=& -2\, e^U D_{IAB} W,\label{flow32}
\end{eqnarray}
where the derivative of the superpotential can be obtained by applying (\ref{DN3}):
\begin{equation}\label{DIABN3}
	D_{IAB} W = \frac{1}{2W} Z_I Z_{AB}.
\end{equation}
The flow equations (\ref{flow31}) and (\ref{flow32}) have been derived previously in \cite{Andrianopoli:2007gt,Ferrara:2008ap}, where also the correct superpotential (\ref{WN3}) has been identified, though using a different approach.

Note that the explicit expression of $D_{IAB}W$ implies right away that only $2n_V$ scalars flow rather than $3n_V$, because
\begin{equation}
	V^A P'_{IAB} \sim V^A D_{IAB} W \sim Z_I V^A Z_{AB} = 0.
\end{equation}

Once the flow equations have been fixed we can provide the identification of the superpotential with the ADM mass by the BPS rewriting of the lagrangian (\ref{LN3}).
The first thing to note is that, using (\ref{DN3}), the black hole potential can be rewritten as a squared expression in terms of the superpotential
\begin{equation}
	V_{BH} = 4 \, \left(\frac12 D_{IAB} W D^{IAB} W\right) + W^2,
\end{equation}
which mimics what happens in $N=2$ in terms of the absolute value of the central charge. 
The action then vanishes on the BPS solutions, up to a boundary term, which is identified with the ADM mass
\begin{equation}
	{\cal L} = \left[U' + e^U W\right]^2 + \frac12 \left(P_{IAB}^\prime + 2\, e^U D_{IAB} W\right)  \left(P^{\prime IAB} + 2\, e^U D^{IAB} W\right) - \left[2 \,e^U W\right]^\prime.
\end{equation}

\subsection{ADM mass constraint} % (fold)
\label{sub:adm_mass_constraint}

Once identified $W$ with the ADM mass, we can prove that it satisfies the relation
\begin{equation}\label{finalN3}
	\frac12 P^C{}_A P^D{}_B \, D_{ICD} D^{IAB} \left(W^2\right) = 4 \left(\frac12\, D_{IAB}W D^{IAB}W\right) + n_V W^2.
\end{equation}

As explained in the introduction, it is crucial to project the second derivatives of the superpotential on the set of scalars active on the black hole solution, otherwise additional terms appear on the right hand side of the equation.
The reason for this has to do with the fact that even if the only derivatives of the superpotential different from zero are along the directions of the running scalars, the second derivative may contain non-zero contributions from orthogonal directions because of the connection terms.
While this projection may seem ad hoc, we stress that this is precisely what we should expect if we want to interpret such relation as a scalar WGC constraint.
Only the scalar mediating the interaction between the black holes should be taken into account.

The derivation is rather easy once one applies the derivatives correctly and uses their properties:
\begin{eqnarray}
	\frac12 P^C{}_A P^D{}_B \, D_{ICD} D^{IAB} \left(W^2\right) &=& \frac14 \, P^C{}_A P^D{}_B \, D_{ICD} D^{IAB} \left(Z_{EF} Z^{EF}\right) \label{deriva1}\\[2mm]
	&=& \frac14 \, P^C{}_A P^D{}_B \, D_{ICD} \left(Z^{EF} D^{IAB}Z_{EF}\right) \label{deriva2}\\[2mm]
	&=& \frac12 \, P^C{}_A P^D{}_B \, D_{ICD} \left(Z^{AB} Z^{I}\right) \label{deriva3}\\[2mm]
	&=& \frac12 \, P^C{}_A P^D{}_B \, \left(n_V Z_{CD}Z^{AB} + Z_I Z^{I} \,2\,\delta^{AB}_{CD}\right) \label{deriva4}\\[2mm]
	&=& n_V W^2 + 2 Z_I Z^I.
\end{eqnarray}
From (\ref{deriva1}) to (\ref{deriva4}) we just use the definition of $W$ and the derivative relations (\ref{DN3}).
The last equality uses once more the definition of $W$ and the fact that $P^A{}_A = 2$.
Finally we recover (\ref{finalN3}) by using (\ref{DIABN3}).

% subsection adm_mass_constraint (end)

% section bps_black_holes_in_n_3_supergravity (end)

\section{BPS black holes in $N=4$ supergravity} % (fold)
\label{sec:bps_black_holes_in_n_4_supergravity}

In the case of $N=4$ supergravity \cite{Bergshoeff:1985ms,Schon:2006kz}, the scalar manifold is factorized and we need to introduce two different types of complex vielbeins, $P_p$ and $P_{IAB}$.
They are in one to one correspondence to the first and second factor in 
\begin{equation}
	{\cal M}_{scalar} = \frac{{\rm SU}(1,1)}{{\rm U}(1)} \times \frac{{\rm SO}(6,n_V)}{{\rm SO}(6) \times {\rm SO}(n_V)}.
\end{equation}
Using the same ansatz for the metric, scalars and vector fields as in the $N=2$ and $N=3$ cases, we can write the reduced 1-dimensional lagrangian as
\begin{equation}
	{\cal L} = \frac14 \, P'_{IAB}P^{\prime IAB} +P_p P_{\bar p}+ (U')^2 + e^{2U} V_{BH},
\end{equation}
where once more \cite{Andrianopoli:1996ve}
\begin{equation}
	V_{BH} = \frac12\, Z_{AB}Z^{AB} + Z_I Z^I.
\end{equation}
Note that in this case the kinetic term of the vector multiplet scalars has an additional 1/2 factor to take into account the redundancy in the representation with the $P_{IAB}$ vielbeins, which indeed satisfy a complex self-duality constraint.

The BPS equations for such theory are
\begin{eqnarray}
	&& \varepsilon_{A}' - \frac{i}{2}\, e^U\, \gamma^0 Z_{AB} \varepsilon^B = 0, \label{susy1N4}\\[2mm]
	&& U' \varepsilon_{A} -i \, e^U\,\gamma^0 Z_{AB} \varepsilon^B = 0, \\[2mm]
	&& P^\prime_p \varepsilon^A = -\frac{i}{2}\, e^U\, \epsilon^{ABCD} Z_{BC} \gamma^0 \varepsilon_D, \\[2mm]
	&& P^\prime_{IAB} \varepsilon^B = i\, e^U \, Z_I \gamma^0 \varepsilon_A, \label{susy4N4}
\end{eqnarray}
and the resulting configurations should preserve 1/4 of the original supersymmetry.
As in the previous case we would like to obtain such configurations by means of two projectors, one that reduces supersymmetries by half and projects on the subspace determined by the highest eigenvalue of the central charge and another one that further reduces supersymmetry by half, relating the projections on the SU(4) indices and on the spinor indices.
The $N=4$ central charge can be skew-diagonalized, so that the squared matrix $M^A{}_B = Z^{AC} Z_{BC}$ has two distinct eigenvalues $e_1$ and $e_2$ with multiplicity 2.
If we assume that $e_1 > e_2 \geq 0$, the ADM mass of the black hole should be identified with $\sqrt{e_1}$ \cite{Andrianopoli:1996ve}.
We therefore want to construct the BPS flow equations as gradient flow equations deriving from a superpotential that coincides with this eigenvalue.
In order to do so, we employ the same technique we employed in the $N=3$ case and construct a projector $P_-^A{}_B$ that projects along the $e_1$ eigenspace and define the superpotential as in (\ref{W}):
\begin{equation}
	W = \sqrt{\frac12\, P_-^C{}_A P_-^D{}_B Z_{CD} Z^{AB}}.
\end{equation}
The projectors can be easily constructed following Schwinger's procedure as
\begin{equation}
	P_1^A{}_B = \frac{Z^{AC}Z_{BC} - e_2\, \delta^A_B}{e_1 - e_2}, \qquad P_2^A{}_B = \frac{Z^{AC}Z_{BC} - e_1\, \delta^A_B}{e_2 - e_1}.
\end{equation}
In order to have a covariant expression in terms of the central charges, we note that we can write the following combinations:
\begin{eqnarray}
	e_1+e_2 &=& \frac12\, Z^{AB}Z_{AB}, \\[2mm]
	(e_1 - e_2)^4 &=& \rm{det} A,
\end{eqnarray}
where
\begin{equation}
	A^A{}_B = 2 Z^{AC} Z_{CB} + \frac12\, \delta^A_B\, Z_{EF}Z^{EF}.
\end{equation}
Hence, after some simple algebra, we see that the projections to the two distinct eigenspaces can be rewritten in terms of 
\begin{equation}
	P_{\pm}^A{}_B = \frac12\left( \delta^A_B \pm \Pi^A{}_B\right),
\end{equation}
where
\begin{equation}
	\Pi^A{}_B = \frac{A^A{}_B}{({\rm det}A)^{1/4}}
\end{equation}
and 
\begin{equation}
	P_- = P_1, \qquad P_+ = P_2.
\end{equation}
Note that $\Pi^2 = {\mathbb 1}_4$, as expected for a projector and therefore we also have the identities
\begin{equation}
	A^2 = \sqrt{\det\,A} \, {\mathbb 1}_4 = \left[Z^{AB} Z_{BC} Z^{CD} Z_{DA} - \frac14\, \left(Z_{EF}Z^{EF}\right)^2\right] {\mathbb 1}_4.
\end{equation}
It is also interesting to note that in this case the projector on the central charge satisfies
\begin{equation}
	P_-^A{}_C P_-^B{}_D Z^{CD} = P_-^A{}_C Z^{CB} = - P_-^B{}_C Z^{CA},
\end{equation}
as follows from the algebraic identities
\begin{eqnarray}
	&&\Pi^A{}_C \Pi^{B}{}_D Z^{CD} = Z^{AB},\\[2mm]
	&&\Pi^A{}_C Z^{CB} = - \Pi^B{}_C Z^{CA}.
\end{eqnarray}
Using this notation, the superpotential can also be expressed as
\begin{equation}\label{supot}
	W = \frac12\, \sqrt{Z_{AB} Z^{AB} + 2 \,(\rm{det} \, A)^{1/4}},
\end{equation}
which can be better handled to compute its derivatives.

Before dealing with the BPS equations we give here the outcome of the application of the covariant derivatives on the superpotential, which can be computed directly by using (\ref{DN4}) on (\ref{supot}):
\begin{eqnarray}
	D_p W &=& \frac{1}{4 W}\, {\rm Pf} \,\bar Z, \label{Dp}\\[2mm]
	D_{IAB} W &=& \frac{1}{2W}\, Z_I Z_{AC} P_-^C{}_B + \frac{1}{4W}\, \delta_{IJ}\, Z^J\, \epsilon_{ABCD} Z^{CF} P_-^D{}_F,\label{DIABW}
\end{eqnarray}
where we introduced the shorthand notation
\begin{equation}
	{\rm Pf}\, Z = \frac14\, \epsilon^{ABCD} Z_{AB} Z_{CD}
\end{equation}
for the Pfaffian of the matrix $Z$.

The BPS flow equations can be obtained from (\ref{susy1N4})--(\ref{susy4N4}) by employing the projectors
\begin{eqnarray}
	&&P^A_+{}_B \varepsilon^B = 0, \\[2mm]
	&& i \gamma^0 \varepsilon^A = \frac{1}{W} Z^{AB} \varepsilon_B.
\end{eqnarray}
The first projector halves the supersymmetries leaving only the spinors in the eigenspace of the maximum eigenvalue of $Z$, while the second further reduces by half the supersymmetries relating different spinor components between them.
We can check consistency of the two projections noting that the first implies 
\begin{equation}
	A^A{}_B \varepsilon^B = - ({\rm det}\, A)^{1/4} \varepsilon^A = \left(-2 W^2 + \frac12\, Z_{CD}Z^{CD}\right) \varepsilon^A
\end{equation}
and therefore
\begin{equation}
	Z^{AB} Z_{BC} \varepsilon^C = - W^2 \, \varepsilon^A,
\end{equation}
while
\begin{equation}
	(\gamma^0)^2 \varepsilon^A = - \frac{i}{W}\, Z^{AB} \gamma^0 \varepsilon_B = \frac{1}{W^2} Z^{AB} Z_{BC} \varepsilon^C = - \varepsilon^A,
\end{equation}
by using the first projection.

Once we use the projectors in the BPS equations we get
\begin{eqnarray}
	&& U' = - e^U\, W, \label{flow41}\\[2mm]
	&& P_p^\prime = -2 \, e^U \, D_{\bar p} W, \\[2mm]
	&& P_{IAB}^\prime = - 2\, e^U\, D_{IAB} W. \label{flow42}
\end{eqnarray}
These flow equations (\ref{flow41})--(\ref{flow42}) have also been discussed in \cite{Andrianopoli:2007gt,Ferrara:2008ap}, together with the superpotential (\ref{supot}), though for the case where only the gravity multiplet is present.

Note that out of the $6n_V$ scalars in $P_{IAB}$, only $2 n_V$ flow, because
\begin{equation}
	P_-^C{}_A P_+^D{}_B \, D_{ICD} W = 0,
\end{equation}
which gives $4 n_V$ conditions.
This follows from (\ref{DIABW}), noting that the first term is fully projected on the $P_-$ subspace and the second is fully projected on the $P_+$ subspace and $P_-^A{}_B P_+^B{}_C = 0$.

The BPS squaring of the action follows by recognizing that
\begin{equation}
	4 ( D_p W D_{\bar p}W + \frac14\, D_{IAB} W D^{IAB} W) = \frac{1}{4W^2} |{\rm Pf} Z|^2 + Z_I Z^I
\end{equation}
and
\begin{equation}
	W^2 + \frac{1}{4W^2} |{\rm Pf} Z|^2 = \frac12 Z_{AB} Z^{AB},
\end{equation}
so that
\begin{equation}\label{VBHN4}
	V_{BH} = 4 ( D_p W D_{\bar p}W + \frac14\, D_{IAB} W D^{IAB} W) + W^2.
\end{equation}
Plugging this into the Lagrangian we eventually obtain
\begin{equation}
	\begin{array}{rl}
	{\cal L} = & \left(U^\prime + e^U \, W\right)^2 + |P_p^\prime+ 2 e^U\, D_{\bar p}W|^2\\[2mm]
	&\displaystyle  + \frac14\left(P_{IAB}^\prime + 2\, e^U\, D_{IAB}W\right)\left(P^{\prime IAB} + 2\, e^U\, D^{IAB}W\right) - (2 e^U W)^\prime,
	\end{array}
\end{equation}
so that again we identify $W$ with the ADM mass.

\subsection{ADM mass constraint} % (fold)
\label{sub:adm_mass_constraint}

The superpotential satisfies an interesting relation, which is the $N=4$ instance of the general expression (\ref{secondversion}):
\begin{equation}\label{D2resultN4}
	\begin{array}{l}
		\displaystyle D_pD_{\bar p}(W^2) + \frac14 \, \left(P_-^A{}_C P_-^B{}_D + P_+^A{}_C P_+^B{}_D\right) D_{IAB} D^{ICD} (W^2) = \\[2mm]
		\displaystyle 4 \left(\frac14\, D_{IAB} W D^{IAB} W\right) + (n_V+1) W^2.
	\end{array}
\end{equation}
Also in this case it is crucial to project on the subspace of complex scalars flowing, given by the $++$ and $--$ combinations of the projectors.

Before beginning with the actual derivation, we note two identities:
\begin{eqnarray}
	&& \epsilon^{ABCD} \Pi_A^{[E} \Pi_B^{F]} = \epsilon^{ABEF} \Pi_A^{[C} \Pi_B^{D]}, \\[2mm]
	&& \epsilon^{ABCD} P_+^{[E}{}_A P_+^{F]}{}_B P_-^G{}_D = \epsilon^{EFBD} P_-^C{}_B P_-^G{}_D.
\end{eqnarray}
We then compute
\begin{equation}
	\begin{array}{l}
	\displaystyle \frac14 \, \left(P_-^A{}_C P_-^B{}_D + P_+^A{}_C P_+^B{}_D\right) D_{IAB} D^{ICD} (W^2) =\\[3mm]
	\displaystyle = \frac14 \, \left(P_-^A{}_C P_-^B{}_D + P_+^A{}_C P_+^B{}_D\right) D_{IAB} \left(Z^I Z^{CE}P_-^D{}_E + \frac12\, \delta^{IJ} \ Z_J\,  \epsilon^{CDEF} Z_{EG} P_-^G{}_F\right)\\[3mm]
	\displaystyle = \frac14 \, \left(P_-^A{}_C P_-^B{}_D + P_+^A{}_C P_+^B{}_D\right) \left(n_V Z_{AB} Z^{CE}P_-^D{}_E + 2 \delta_{AB}^{CE} \, Z_I Z^I P_-^D{}_E - \frac12 Z^I Z^{CE} D_{IAB} \Pi^D{}_E \right. \\[3mm]
	 \displaystyle+ \frac14 \, n_V\, \epsilon_{ABPQ} Z^{PQ} \epsilon^{CDEF} Z_{EG} P_-^G{}_F + \frac12\, Z_I Z^I \epsilon_{ABEG} \epsilon^{CDEF} P_-^G{}_F \\[3mm]
	 \left. \displaystyle - \frac14 \delta^{IJ} \ Z_J\,  \epsilon^{CDEF} Z_{EG} D_{IAB}\Pi^G{}_F\right).
	\end{array}
\end{equation}
Using projector identities like $P_-^2 = P_-$, $P_+ P_- = 0$, $\epsilon^{ABCD} P_-^{[E}{}_A P_-^{F]}{}_B P_-^G{}_D = 0$ and $\Pi^A{}_B D_{IEF} \Pi^B{}_C = 0$, we see that 
\begin{equation}\label{relaN4}
	\begin{array}{l}
	\displaystyle \frac14 \, \left(P_-^A{}_C P_-^B{}_D + P_+^A{}_C P_+^B{}_D\right) D_{IAB} D^{ICD} (W^2) =\\[3mm]
	\displaystyle = n_V W^2 + Z_I Z^I - \frac14\, Z_I Z^I\, \left(D_{IAB} \Pi^B{}_C - \Pi^E{}_A \Pi^B{}_C D_{IEF} \Pi^F{}_B\right) Z^{AC}\\[2mm]
	\displaystyle - \frac18 \delta^{IJ} Z_J\, \epsilon^{ABCD} Z_{CG} \left(D_{IAB} \Pi^G{}_D + \Pi^E{}_A \Pi^F{}_B\, D_{IEF} \Pi^G{}_D\right).
	\end{array}
\end{equation}
Now, recalling that $Z^{AC} = Z^{EF} \Pi^A{}_E \Pi^C{}_F$, the third term vanishes, and we can see that also the last one vanishing upon using the identities above: 
\begin{equation}
	\begin{array}{l}
	\displaystyle - \frac18 \,\delta^{IJ} \,Z_J\, \epsilon^{ABCD} Z_{CG} \left(D_{IAB} \Pi^G{}_D + \Pi^E{}_A \Pi^F{}_B\, D_{IEF} \Pi^G{}_D\right)\\[2mm]
	\displaystyle =-\frac14Z_J Z_{CG} D^{JCD} \Pi^G{}_D - \frac18\, \delta^{IJ} Z_J\, \epsilon^{ABCD} Z_{CG} \Pi^E{}_A \Pi^F{}_B D_{IEF} \Pi^G{}_D\\[2mm]
	\displaystyle =-\frac14\, Z_J \left( Z_{AG} - Z_{EF} \Pi^E{}_A \Pi^F{}_G \right) D^{JAB} \Pi^G{}_B = 0.\\[2mm]
	\end{array}
\end{equation}
Finally, we use (\ref{VBHN4}) in (\ref{relaN4}) to recover (\ref{D2resultN4}).

% subsection adm_mass_constraint (end)
% section bps_black_holes_in_n_4_supergravity (end)

\section{Comments} % (fold)
\label{sec:comments}

In previous sections we have built evidence that for asymptotically flat BPS black holes in 4 dimensions we have a differential constraint on their ADM mass of the form
\begin{equation}
	P^a{}_b D_a \overline{D}^b (M^2) = 4 D_a M \overline{D}^a M + n\, M^2,
\end{equation}
where derivatives are taken only with respect to the running complex scalars.
Starting from this result, we can now use the WGC to obtain a general constraint on the scalar-dependent masses of the various fields.
For a generic charged black hole in the presence of scalar fields we have that
\begin{equation}\label{BHrel}
	M^2 + \Sigma^2 - Q_{\infty}^2 \geq 0,
\end{equation}
where $M$ is the ADM mass of the black hole, $\Sigma$ represents the scalar charges and $Q_{\infty}$ are the U(1) charges at infinity.
Our relation can also be written as
\begin{equation}
	D^2 M^2 = n\, M^2 + \Sigma^2 = (n -1)M^2 + (M^2 + \Sigma^2),
\end{equation}
where $\Sigma = 2 D W = 2 DM$.
Using the black hole relation (\ref{BHrel}) we therefore find
\begin{equation}
	M^2 + \Sigma^2 - Q_{\infty}^2  = n\, M^2 + \Sigma^2 - D^2 M^2 \geq 0, 
\end{equation}
which implies that the particle needed to discharge the black hole should satisfy the opposite inequality
\begin{equation}
	D^2 m^2(\phi) \geq n\, m(\phi)^2 + 4 (D m(\phi))^2.
\end{equation}

This is a rather strong constraint on the possible moduli dependence of the masses of particles in effective theories.
While we would like to take such relation and use it as a novel scalar WGC, we should first inspect it more closely to better understand its requirements and limits.

First of all we would like to point out that it is difficult to extract a simple universal behaviour of the masses as a function of the scalar fields.
Take for instance conjugate BPS configurations in the $N=2$ STU model
\begin{eqnarray}
	K &=& - \log\left[i (s-\bar s)(t - \bar t)(u - \bar u)\right], \\[2mm]
	Z_1 &=& e^{K/2}\left(p^0 stu -q_1 s - q_2 t - q_3 u\right), \\[2mm]
	Z_2 &=& e^{K/2}\left(-q^0 stu +p_1 tu +p_2 su +p_3 st\right).
\end{eqnarray}
Using a real parameterization
\begin{equation}
	s = \frac{\sigma}{M} + i\, e^{-\sqrt2\phi_s/M}, \qquad 	t = \frac{\tau}{M} + i\, e^{-\sqrt2\phi_t/M}, \qquad 	u = \frac{\nu}{M} + i\, e^{-\sqrt2\phi_u/M}, 
\end{equation}
we see that only the $\phi_{s,t,u}$ scalars flow along the black hole solution and the ADM mass $M_{ADM} = |Z|$ has a very simple and yet different dependence on them, namely
\begin{equation}
	\begin{array}{rl}
	M_{ADM} \sim &-p^0 e^{-(\phi_s+\phi_t+\phi_u)/(\sqrt2 M)} + q_1 e^{(-\phi_s+\phi_t+\phi_u)/(\sqrt2 M)}  \\[2mm]
	&+ q_2 e^{(\phi_s-\phi_t+\phi_u)/(\sqrt2 M)}+ q_3 e^{(\phi_s+\phi_t-\phi_u)/(\sqrt2 M)},
	\end{array}
\end{equation}
for the $Z_1$ charge and
\begin{equation}
	\begin{array}{rl}
	M_{ADM} \sim &-q_0 e^{(\phi_s+\phi_t+\phi_u)/(\sqrt2 M)} + p^1 e^{(\phi_s-\phi_t-\phi_u)/(\sqrt2 M)} \\[2mm]
	&+ p^2 e^{(-\phi_s+\phi_t-\phi_u)/(\sqrt2 M)}+ p^3 e^{(-\phi_s-\phi_t+\phi_u)/(\sqrt2 M)}
	\end{array}
\end{equation}
for the $Z_2$ charge.
We can interpret the resulting behaviour as the outcome of the sum over different states, whose masses either exponentially vanish or blow-up in the $\phi_{s,t,u} \to \pm \infty$ limit towards the boundary of the moduli space.
This is the expected behaviour to be compatible with the swampland distance conjecture and also with the microscopic interpretation of the black hole charges with D-branes wrapping cycles of the internal manifold (in this case D0, D2, D4 and D6-branes on 0, 2, 4 and 6-cycles of ${\mathbb T}^6/({\mathbb Z}_2 \times {\mathbb Z}_2)$).

Another aspect that emerges from this analysis is that it is crucial in the relation to have a second covariant derivative spanning over all active \emph{complex} scalars.
In the $N=2$ example that we just presented, $\sigma = 0 = \tau = \nu$ along the whole solution \cite{Behrndt:1996hu}, but the identity is fulfilled only if the terms $g^{\sigma \sigma} \partial_{\sigma}^2 |Z|$ and $g^{\sigma \sigma} \gamma^{\phi_s}_{\sigma \sigma}\partial_{\phi_s} |Z|$ are taken into account (and their analogous terms for the $t$ and $u$ fields).
Without considering these terms one would not obtain a differential equation on $M_{ADM}$ resulting in the expected behaviour in $\phi_{s,t,u}$.
This clearly hampers the possibility of a straightforward generalization to theories where the moduli fields do not come in complex form.

The last point that is quite peculiar of this relation is that its validity rests on the \emph{sum} over all complex scalars contributing to the BPS configuration.
This means that we are not able at this stage to extract a strong form of the inequality, to be valid for \emph{each} scalar, like the one proposed in \cite{Gonzalo:2019gjp}.

While the formula we proposed for the differential relation on the ADM mass of a BPS black hole has been written in a form that is independent of the number of supersymmetries, we should stress that we completed the proof only for $N=2$, $N=3$ and $N=4$.
We do not foresee obstacles to a further extension to $N=5$, $N=6$ or $N=8$, and in fact, in \cite{Andrianopoli:2007gt} one can find the identification of the superpotential with the appropriate eigenvalue of the central charge matrix, but it is clear that technically computations become much more involved because the projectors needed have a rather complicated expression in terms of traces and determinants of combinations of the central charges.  
This is an obvious possible extension of the work reported here.

Another possible extension of this work is the analysis of the extremal non-BPS configurations in extended supegravity, along the lines of what done in \cite{Ceresole:2007wx} for the $N=2$ case.
The $N=2$ case has been already been discussed in \cite{Palti:2017elp}, but for $N>2$ one can imagine that different possible superpotentials appear, according to the branch of non-BPS extremal solutions (see \cite{Andrianopoli:2006ub} for an overview of the possibilities).
Some instances of such superpotentials have been discussed in \cite{Ferrara:2008ap} and it would be interesting to see if they all satisfy the same differential constraint.

% section comments (end)

\bigskip

\bigskip

\noindent
\textbf{Acknowledgments}

\noindent
We thank D.~Cassani, A.~Van Proeyen, I.~Valenzuela and especially E.~Palti for very useful comments and discussions.
The work of GD is supported in part by MIUR-PRIN contract 2017CC72MK003.

%%%%%%%%%%%%%%%%%%%%%%%%%%%%%%%%%%%%%%%%%%%%%%%%%%%%%%%%%%%%%%

\end{document}